\documentclass[10pt, conference]{IEEEtran}
\IEEEoverridecommandlockouts

\usepackage{cite}
\usepackage{amsmath,amssymb,amsfonts}
\usepackage{algorithmic}
\usepackage{graphicx}
\usepackage{textcomp}
\usepackage{xcolor}
\def\BibTeX{{\rm B\kern-.05em{\sc i\kern-.025em b}\kern-.08em
    T\kern-.1667em\lower.7ex\hbox{E}\kern-.125emX}}

\usepackage{enumitem}
\usepackage[utf8]{inputenc}
\usepackage[linesnumbered,ruled,vlined,noend]{algorithm2e}
\usepackage{listings}
\usepackage{xcolor}
\usepackage{multirow} 
\usepackage{float}
\usepackage{graphicx}
\usepackage{subcaption}
\usepackage{caption}
\usepackage{makecell}
\usepackage{tcolorbox}
\usepackage{hyperref}
\usepackage{pgfplots}
\usepackage{booktabs} 
\usepackage{balance}
\pgfplotsset{width=8cm,compat=1.9}

\hypersetup{
  colorlinks,
  linkcolor=red, 
  urlcolor=blue,
  citecolor=black,
  bookmarksnumbered,
  unicode
}

\usepackage{cleveref}
\crefname{section}{§}{§§}
\Crefname{section}{§}{§§}

\definecolor{codeblue}{rgb}{0.29, 0.59, 0.82}
\definecolor{codegreen}{rgb}{0,0.6,0}
\definecolor{codered}{rgb}{0.6,0.0,0.0}
\definecolor{codegray}{rgb}{0.5,0.5,0.5}
\definecolor{codepurple}{rgb}{0.58,0,0.82}
\definecolor{backcolour}{rgb}{0.96,0.96,0.96}

\definecolor{codegreenbg}{rgb}{0.9,0.99,0.9}
\definecolor{coderedbg}{rgb}{0.99,0.9,0.9}

\lstdefinestyle{ExampleStyle}{
    commentstyle=\color{codegreen},
    keywordstyle=\color{codepurple},
    numberstyle=\tiny\color{codegray},
    stringstyle=\color{codepurple},
    basicstyle=\tt\footnotesize,
    breakatwhitespace=false,         
    breaklines=true,                 
    captionpos=b,                    
    keepspaces=true,                 
    numbers=left,                    
    numbersep=5pt,                  
    showspaces=false,                
    showstringspaces=false,
    showtabs=false,                  
    tabsize=2,
    frame=single, 
    frameround=tttt,
    rulecolor=\color{black}, 
    framexleftmargin=10pt, 
    framextopmargin=6pt,
    framexbottommargin=6pt, 
    xleftmargin=\parindent,
    xrightmargin=\parindent,
}

\begin{document}

\title{Hybrid Automated Program Repair by Combining Large Language Models and Program Analysis}

\author{%
    \IEEEauthorblockN{Fengjie Li\textsuperscript{1}, Jiajun Jiang\textsuperscript{1}\thanks{$^*$Jiajun Jiang is the corresponding author for this work.}, Jiajun Sun\textsuperscript{1} and Hongyu Zhang\textsuperscript{2}}
    \IEEEauthorblockA{%
    \textsuperscript{1}College of Intelligence and Computing, Tianjin University, China\\
    \textsuperscript{2}School of Big Data and Software Engineering, Chongqing University, China \\
    fengjie@tju.edu.cn, jiangjiajun@tju.edu.cn, sjjtianjin@tju.edu.cn, hyzhang@cqu.edu.cn
    }
}

\newcommand{\toolink}[1]{\href{https://github.com/Feng-Jay/GiantRepair}{https://github.com/Feng-Jay/GiantRepair}}
\newcommand{\TOOL}[1]{\textsc{GiantRepair}}

\newcommand{\jiajun}[1]{\textcolor{blue}{[Jiajun: #1]}}
\newcommand{\hy}[1]{\textcolor{red}{[HY: #1]}}
\newcommand{\fj}[1]{\textcolor{orange}{[FJ: #1]}}

\newcommand{\TODO}[1]{\textcolor{red}{TODO: #1}}

\newcommand{\MODI}[1]{\textcolor{orange}{#1}}

\lstdefinestyle{java}{ 
	language=java,
	basicstyle=\scriptsize\ttfamily, 
	breakatwhitespace=false, 
	breaklines=true, 
	captionpos=b, 
	commentstyle=\color[rgb]{0.0, 0.5, 0.69},
	deletekeywords={}, 
	escapeinside={<@}{@>},
	firstnumber=1, 
	frame=lines, 
	frameround=tttt, 
	keywordstyle={[1]\color{blue!90!black}},
	keywordstyle={[3]\color{red!80!orange}},
	morekeywords={String,int}, 
	numbers=none, 
	numbersep=-8pt, 
	numberstyle=\tiny\color[rgb]{0.1,0.1,0.1}, 
	rulecolor=\color{black}, 
	showstringspaces=false, 
	showtabs=false, 
	stepnumber=1, 
	stringstyle=\color[rgb]{0.58,0,0.82},
	tabsize=2, 
	backgroundcolor=\color{white}
}

\newcommand{\lin}[1]{{\scriptsize \textcolor{darkgray}{#1}}}
\newcommand{\codeIn}[1]{{\ttfamily #1}}

\newcommand{\distance}{1pt}
\setlength{\textfloatsep}{\distance}
\setlength{\floatsep}{\distance}
\setlength{\dbltextfloatsep}{\distance} 
\setlength{\dblfloatsep}{\distance} 

\maketitle

\begin{abstract}
Automated Program Repair (APR) has garnered significant attention 
due to its potential to streamline the bug repair process for human developers. Recently, LLM-based APR methods have shown promise in repairing real-world bugs. However, existing APR methods often utilize patches generated by LLMs without further optimization, resulting in reduced effectiveness due to the lack of program-specific knowledge. Furthermore, the evaluations of these APR methods have typically been conducted under the assumption of perfect fault localization, which may not accurately reflect their real-world effectiveness.
To address these limitations, this paper introduces an innovative APR approach called \TOOL{}. Our approach leverages the insight that LLM-generated patches, although not necessarily correct, offer valuable guidance for the patch generation process. 
Based on this insight, \TOOL{} first constructs patch skeletons from LLM-generated patches to confine the patch space, and then generates high-quality patches tailored to specific programs through context-aware patch generation by instantiating the skeletons. To evaluate the performance of our approach, we conduct two large-scale experiments. The results demonstrate that \TOOL{} not only effectively repairs more bugs (an average of 27.78\% on Defects4J v1.2 and 23.40\% on Defects4J v2.0) than using LLM-generated patches directly, but also outperforms state-of-the-art APR methods by repairing at least 42 and 7 more bugs under perfect and automated fault localization scenarios, respectively.
\end{abstract}





\section{Introduction}
\label{sec:Introduction}

As the scale and complexity of modern software systems continue to grow, the prevalence of software bugs has also risen, resulting in substantial financial and operational loss for organizations and end-users. Addressing these bugs requires a significant investment of time and effort from developers. As a result, 
Automated Program Repair (APR), which seeks to automatically generate correct patches for buggy code, has garnered considerable interest from both academia and industry.

In the past years, numerous APR approaches have been proposed with the goal of enhancing the quality of automatically generated patches and making them more practical for real-world use~\cite{le2011genprog,xuan2016nopol,le2017s3,long2015staged,mechtaev2016angelix,martinez2016astor,Le2016HistoryDP,long2016automatic,xin2017leveraging,xiong2017precise,long2017automatic,hua2018towards, wen2018context,jiang2018shaping,liu2019tbar,liu2019avatar,ghanbari2019practical,jiang2019inferring}. These approaches include generating patches through predefined repair templates~\cite{martinez2016astor,long2017automatic,hua2018towards,ghanbari2019practical,jiang2019inferring,liu2019avatar}, heuristic rules~\cite{le2011genprog, Le2016HistoryDP,long2016automatic,xin2017leveraging,xiong2017precise,wen2018context,jiang2018shaping}, and constraint solving techniques~\cite{xuan2016nopol,le2017s3,long2015staged,mechtaev2016angelix}. While these methods have proven effective in addressing some real-world bugs, the number of correct fixes remains limited~\cite{xia2023automated, xia2023plastic}. The reason is that it is difficult for these methods to tackle the large search space of diverse bugs in real applications. For example, the template-based methods rely on high expertise and manual efforts to construct the templates; the heuristic-based methods are less effective when facing the rapid growing of patch space; while the constraint-based methods suffer from scalability issues. Although deep-learning-based APR methods have significantly improved the repair ability by utilizing the latest advance of deep learning techniques~\cite{chen2019sequencer,lutellier2020coconut,li2020dlfix, zhu2021syntax, jiang2021cure, ye2022neural, zhu2023tare}, many previous studies~\cite{fu2022vulrepair,Xia2022LessTM,feng2024prompting} pointed out that their repair capacity relies on the quality of training data, making them hard to repair bugs that have not encountered during training.

Recently, Large Language Models (LLMs) have demonstrated promising results across various software engineering tasks, e.g., code search~\cite{guo2022unixcoder}, program synthesis~\cite{nijkamp2022codegen}, defect detection~\cite{wang2023codet5+}, code summarization~\cite{ahmed2022few} and so on. Some recent studies~\cite{kolak2022patch, Xia2022LessTM, prenner2022can, jiang2023impact, xia2023automated,xia2023plastic,wei2023copiloting, zhang2023gamma, silva2023repairllama} have also explored the application of LLMs in automated program repair. The initial results demonstrate their ability to correctly repair real-world bugs, including those that were previously unrepairable by existing APR approaches. The promising outcomes suggest the potential of LLMs in developing more effective APR methods.

While recent studies~\cite{xia2023plastic, wei2023copiloting, zhang2023gamma, jiang2023impact, xia2023automated} have explored the use of LLMs for automated program repair, there are still significant limitations that need to be addressed:

\begin{enumerate}
    \item Existing LLM-based APR approaches directly leverage the patches generated by LLMs, without further optimization or refinement. However, LLMs may struggle to generate patches that correctly incorporate program-specific elements like local variables and domain-specific method calls. This means that even if the LLM-generated patches are “close” to the desired solution, they may still fail to pass the test cases. 
    How to effectively utilize these ``incorrect'' patches to improve the overall repair ability remains a largely unexplored question.

    \item  Evaluations of LLM-based APR approaches have so far been conducted under the assumption of perfect fault localization, where the faulty locations are already known. This is an unrealistic scenario, as in practice, automated fault localization techniques are often inaccurate. The real-world performance of LLM-based APR approaches under the more realistic setting of automated fault localization is yet to be thoroughly investigated.
\end{enumerate}

To address these limitations, a more comprehensive and practical evaluation of LLM-based APR approaches is required. This should involve exploring methods to better leverage the insights from ``incorrect'' patches generated by LLMs, as well as assessing the performance of these techniques under the more challenging scenario of automated fault localization. Addressing these limitations is crucial for understanding the true potential and practical applicability of LLM-based approaches in the field of automated program repair.


In this paper, we aim to address these two limitations. Specifically,
to address the first limitation, we propose a novel automated program repair approach, named \TOOL{}. The key insight behind \TOOL{} is that patches generated by LLMs, although not always correct, can still provide valuable guidance for the patch generation process. Specifically, \TOOL{} first leverages LLMs to efficiently generate a diverse set of candidate patches. It then abstracts these candidate patches into a set of \textit{patch skeletons} that capture the core structures of the patches. These patch skeletons are then used to guide the subsequent context-aware patch generation process, where the patches are refined and instantiated to fit the specific program context. This two-step approach has several advantages. First, the use of patch skeleton helps to confine the search space of possible patches, making the patch generation process more efficient and effective. Second, by combining the strengths of LLMs (for initial patch generation) and context-aware refinement (for patch instantiation), \TOOL{} is able to generate high-quality patches that can correctly fix real-world bugs.
To address the second limitation and comprehensively evaluate the performance of \TOOL{}, we assess its effectiveness not only under the assumption of perfect fault localization (as done in previous studies~\cite{xia2023plastic, wei2023copiloting, zhang2023gamma, jiang2023impact, xia2023automated, xia2023keep}), but also with the more realistic scenario of automated fault localization. This allows us to better understand the practical applicability of \TOOL{} in real-world settings.

We have conducted two large-scale experiments using the widely-used Defects4J benchmark~\cite{just2014defects4j} to evaluate \TOOL{} in two different application scenarios.

\begin{enumerate}
    \item In the first scenario, we compared the repair results of individual LLMs with and without integrating \TOOL{}. The results showed that \TOOL{} improved the repair performance of individual LLMs by correctly repairing an average of 27.78\% and 23.40\%  more bugs on Defects4J v1.2 and Defects4J v2.0.

    \item In the second scenario, we integrated \TOOL{} with existing LLMs to form a standalone APR and compared its repair results with existing state-of-the-art APR approaches. Under the assumption of perfect fault localization, \TOOL{} successfully repaired 171 bugs, outperforming the best state-of-the-art APR approaches by repairing at least 42 more bugs. When with the more realistic scenario of automated fault localization, \TOOL{} can still repair at least 7 more bugs than the best-performing APR approaches.
\end{enumerate}

Overall, the experimental results demonstrate the effectiveness and generality of \TOOL{}, providing new insights for future research in the field of APR. The results highlight the potential for better utilization of LLM outputs for improved APR.
To sum up,
this paper makes the following major contributions.
\begin{itemize}
    \item An innovative automated program repair technique that leverages the capabilities of LLMs and context-aware patch refinement.
    \item A novel patch generation method that extracts patch skeleton from LLM-generated patches to confine the patch space for better APR.
    \item A comprehensive evaluation in two application scenarios, and the experimental results confirm the effectiveness and generalizability of our approach.
    \item We have open-sourced our implementations and all experimental data to facilitate future research in this field. \toolink{}
\end{itemize}

\section{Motivating Examples}
\label{sec:motivation}


\newcommand{\hlc}[2]{{\setlength\fboxsep{0pt}\hspace{-3pt}\colorbox{#1} 
		{\begin{minipage}{\dimexpr\columnwidth-1\fboxsep+0pt\relax}
				\codeIn{\strut\hspace{3pt}#2}
			\end{minipage}}}}
   
In this section, we show two real-world examples from our experiment to demonstrate how incorrect LLM-generated patches can be utilized to guide the patch generation process, thereby motivating the need for our context-aware patch generation method.
Listing~\ref{lst:JacksonDatabind-51} shows the developer patch and LLM's patch for the bug JacksonDatabind-15 from Defects4J~\cite{just2014defects4j}. In this paper, a line of code starting with ``+'' denotes a newly added line while lines starting with ``-'' denote lines to be deleted. 

\begin{lstlisting}[style=java, numbers=none,label=lst:JacksonDatabind-51,caption=Patches of JacksonDatabind-51 from Defects4J]
// Developer patch
<@\hlc{green!15}{+ if (!type.hasGenericTypes()) \{} @>
      type = ctxt.getTypeFactory()
        .constructSpecializedType(_baseType, 
            type.getRawClass());
<@\hlc{green!15}{+ \} } @>
// LLM's patch
<@\hlc{yellow!15}{+ if (!type.equals(\_baseType)) \{} @>
      type = ctxt.getTypeFactory()
        .constructSpecializedType(_baseType, 
            type.getRawClass());
<@\hlc{yellow!15}{+ \}} @>
\end{lstlisting}

The developer patch indicates the need to add an \codeIn{if} condition to fix the bug. However, accurately addressing this bug presents several challenges for existing APR methods: (1) pinpointing the exact faulty line of code is difficult; (2) determining the necessity of introducing a new \codeIn{if} statement is complex; (3) the specific conditional expression (\codeIn{!type.hasGenericTypes()}) is domain-specific and may not be applicable elsewhere. These challenges result in a vast search space for potential patches. Consequently, none of the existing approaches in our experiment, including the latest LLM-based APRs, were able to successfully address this issue.

Upon comparing the LLM’s patch with the developer patch, it becomes evident that they are somewhat similar -- both introduce a new \codeIn{if} statement with a method call as the condition. Although the LLM’s patch lacks domain-specific knowledge (i.e., the method call \codeIn{hasGenericTypes()}), it still offers valuable guidance by providing a similar patch structure (e.g., the \codeIn{if} statement), effectively narrowing down the search space for potential patches. However, effectively leveraging these LLM-generated patches for improving APR remains challenging due to the diverse and complex nature of real-world situations. For example, in Listing~\ref{lst:Math-10}, another real-world bug repair example is presented. Here, the desired patch involves inserting a new \codeIn{Assignment} for the variable \codeIn{result[resultOffset]}, while the LLM’s patch introduces a new \codeIn{for} loop statement. In this case, only a portion of the LLM’s patch may be useful, as reusing the entire \codeIn{for} loop will not pass the test cases. Additionally, updating the indices of the array accesses to variables \codeIn{x} and \codeIn{y} is also necessary to construct the correct patch.

\begin{lstlisting}[style=java, numbers=none,label=lst:Math-10,caption=Patches of Math-10 from Defects4J]
// Developer patch
<@\hlc{green!15}{+ result[resultOffset] = } @>
<@\hlc{green!15}{+ \qquad    FastMath.atan2(y[yOffset], x[xOffset]);} @>
// LLM's patch
<@\hlc{yellow!15}{+ for (int i = 0; i < tmp1.length; ++i) \{ } @>
<@\hlc{yellow!15}{+ \qquad result[resultOffset + i] =} @>
<@\hlc{yellow!15}{+  \qquad\qquad  FastMath.atan2(y[yOffset + i],x[xOffset + i]);}@>
<@\hlc{yellow!15}{+ \}} @>
\end{lstlisting}

To address the challenges outlined above, this paper introduces a context-aware and adaptive patch generation method aimed at effectively reusing patches generated by LLMs. As previously discussed, LLM’s patches offer valuable patch structures. Therefore, the core concept of our approach is to construct patch skeletons from LLM’s patches to limit the patch space, and then generate high-quality patches through context-aware skeleton instantiation using static analysis. This enables the generation of patches tailored to specific programs.

\section{Approach}
\label{sec:Approach}

\begin{figure*}[htbp]
    \centering
    \includegraphics[width=\textwidth]{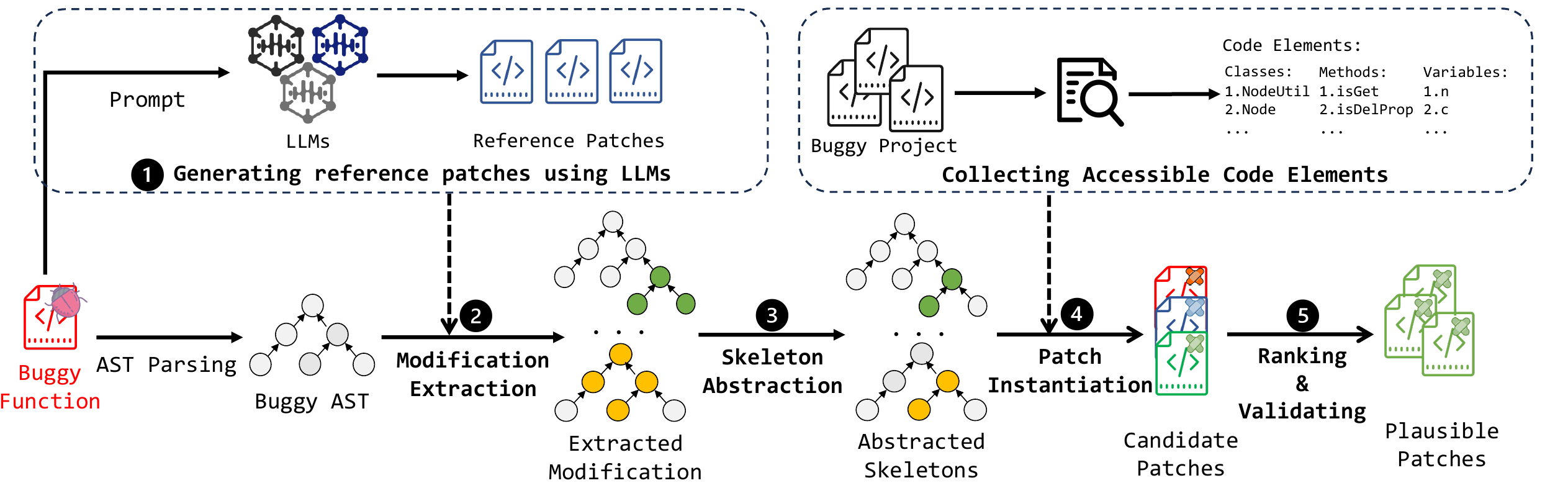}
    \caption{Overview of \TOOL{}}
    \label{fig:overview}
\end{figure*}

In this section, we provide a detailed explanation of our approach, i.e., \TOOL{}. As previously mentioned, our approach is based on the insight that LLM-generated patches, while not always correct, can offer valuable guidance on patch structure for constraining the patch space. Therefore, our patch generation process consists of two key components: \textbf{skeleton construction} and \textbf{patch instantiation}.
(1) The skeleton construction component (Section~\ref{sec:section3.1}) involves extracting a set of code modifications from the provided LLM-generated patches by comparing the buggy code and patched code through tree-level differencing over the abstract syntax tree. These modifications are then abstracted into patch skeletons using a set of abstraction rules.
(2) The patch instantiation component (Section~\ref{sec:section3.2}) takes these skeletons and instantiates them using valid (e.g., defined under certain contexts) and compatible (e.g., satisfying type constraints) program elements through static analysis, resulting in executable patches.
Finally, \TOOL{} evaluates the correctness of the patches by running test cases according to a patch ranking strategy (Section~\ref{sec:section3.3}). Figure~\ref{fig:overview} provides an overview of our approach, and in the following sections, we will delve into each step in detail.

\subsection{Skeleton Construction}
\label{sec:section3.1}

As mentioned in Section~\ref{sec:motivation}, it is often the case that not all the code changes in an LLM-generated patch are desirable under different contexts. Therefore, it is necessary to disassemble the code changes in a patch into individual components, allowing them to be applied independently. To achieve this, \TOOL{} incorporates a modification extraction process that aims to identify all concrete code changes through tree-level code differencing between the buggy code and the patched code. Each identified modification then will be abstracted into a patch skeleton for subsequent patch generation.


\subsubsection{Modification Extraction}
\label{sec:section3.1.1}

In order to extract concrete modifications, \TOOL{} performs a tree-level code matching and differencing process. Specifically, \TOOL{} endeavors to match the code elements from the buggy code and the patched code and then generates code modifications for those elements that are different between these two sides. In particular, we consider statement-level code matching rather than the expression-level. The reasons are twofold: (1) matching statements are more efficient and less likely to produce incorrect matching because different statements tend to be diverse while the finer-grained expressions have a larger possibility to be the same at different locations; (2) the search space of statement-level code changes is relatively small and should be 
efficient for skeleton construction and instantiation. 

\begin{algorithm}
\caption{AST Node Matching}
\label{algo:matching}
\SetAlgoLined
\SetKwProg{Fn}{func}{:}{}
\SetKwFunction{Fmatch}{match}
\SetKwFunction{FmatchNode}{matchNode}
\Fn{\Fmatch{$a$\text{: ASTNode}, $b$\text{: ASTNode}}}{
\textit{mapping} $\leftarrow$ \FmatchNode{a, b}\\
\textit{mapping.sortBySimilarity()}\\
\textit{result} $\leftarrow \emptyset$ , \textit{dups} $\leftarrow \emptyset$ \\
\ForEach{$\langle a,b\rangle \in$ \textit{mapping}}{
	\If{a $\notin$ dups and b $\notin$ dups}{
		\textit{dups} $\leftarrow$ \textit{dups} $\cup \{ a, b \}$\\
		\textit{result} $\leftarrow$ \textit{result} $\cup \{\langle a,b\rangle \}$ 
	}
	
}
\KwRet{$result$}
}

\Fn{\FmatchNode{$a$\text{: ASTNode}, $b$\text{: ASTNode}}}{
\textit{result} $\leftarrow \emptyset$\\
\If{type(a) = type(b)}{
\textit{result} $ \leftarrow$ \textit{result} $\cup \{\langle a,b \rangle \}$\\
\tcc{{\footnotesize$\times$: Cartesian product of two sets}}
\ForEach{$\langle a',b'\rangle \in$ \textit{children(a)} $\times$ \textit{children(b)}}{
		\textit{result} $\leftarrow$ \textit{result} $\cup$ \FmatchNode{$a'$,$b'$}\\
	}
}
\KwRet{result}
}
\end{algorithm}

Algorithm~\ref{algo:matching} presents the matching algorithm in \TOOL{}. In particular, we use the function \textit{type(a)} to represent the AST node type  (e.g., IfStatement) of node \textit{a}, and use
\textit{children(a)} to return a set of statements that are child nodes of \textit{a}.
In general, it takes two AST nodes (one from the buggy code and one from the patched code) as inputs and recursively performs a top-down matching process following a greedy strategy -- two AST nodes (i.e., statements) can match each other as long as their node types are the same (line 12). After this process, one statement from the buggy code may have more than one matched statement from the patched code (according to line 14). To obtain the best matching and extract the fewest modifications, \TOOL{} removes the redundant matches and only preserves the best one by measuring the similarity between the matched statements (line 3). Formula~\ref{eq:sim} defines the computation of similarity between two statements. In the formula, we use ``\textit{atomic stmts}'' to denote those statements that cannot be decomposed into finer-grained ones, e.g., ExpressionStatement, while use ``\textit{ensembled stmts}'' to represent those that are ensembled by other statements, e.g., IfStatement. In addition, the function \textit{editDistance(a,b)} computes the token-level edit distance~\cite{ke2015repairing,koyuncu2020fixminer} between code snippets corresponding to node \textit{a} and \textit{b}, while $C_a$ denotes the child statements of $a$.
\begin{equation}\label{eq:sim}
sim(a,b) =
\begin{cases}
\frac{\textit{editDistance}(a, b)}{\textit{length}(a)} &\textit{:atomic stmts.} \\
\frac{1}{|C_a|}\sum\limits_{c_1 \in C_a} \mathop{\max}\limits_{c_2 \in C_b } \textit{sim}(c_1, c_2) &\textit{:ensembled stmts.}\\
0 & \textit{:otherwise} \\
\end{cases}    
\end{equation}
Once obtaining the matching results from Algorithm~\ref{algo:matching}, \TOOL{} extracts the concrete modifications at the statement level. Suppose that statements \textit{a} and \textit{b} are from the buggy and the patched code respectively. Then, \TOOL{} may generate the following modifications according to the matching results:
\begin{description}
    \item[Update(a,b)]: replaces statement \textit{a} with statement \textit{b} if \textit{a} matches \textit{b} but they are not completely the same code;
    \item[Insert(b, i)]: inserts the statement \textit{b} into the buggy code as the $i_{th}$ child statement of \textit{p'} if \textit{b} does not match any statement but its parent node \textit{p} matches statement \textit{p'}. \textit{i} is the index of \textit{b} in \textit{p};
    \item[Delete(a)]: deletes the statement \textit{a} from the buggy code;
\end{description}

Intuitively, after applying all extracted modifications to the buggy code, it will be transformed into the same one as the patched code.

\subsubsection{Skeleton Abstraction}
\label{sec:section3.1.2}


As aforementioned, the modifications extracted from LLM-generated patches contain valuable guidance. Specifically, they often make changes at the correct locations and possess AST structures similar to the correct fixes. However, directly applying these modifications may not necessarily produce correct patches as they may introduce incorrect program elements that are not applicable to certain contexts under repair. To overcome this issue, \TOOL{} involves a code abstraction process.
That is, after obtaining the concrete modifications according to the algorithm explained above, \TOOL{} performs code abstraction that constructs patch skeletons by removing concrete program elements while preserving the code structure for confining the subsequent patch generation. Specifically, \textit{\TOOL{} performs the abstraction process only for the statement \textit{b} appearing in modifications \textbf{Update(a,b)} and \textbf{Insert(b,i)}}. In contrast, for modification \textbf{Delete(a)}, abstraction is not necessary as it will not introduce any new elements to the program.

\newcommand{\highlight}[1]{\textcolor{red}{#1}}
\newcommand{\keyw}[1]{\textcolor{blue}{#1}}
\begin{table*}[tbp]
    \caption{Abstraction rules for patch skeleton construction regarding different AST node types. The \textcolor{blue}{blue} tokens in the skeleton are keywords that will be \textit{extended} from the source code during abstraction, while the \textcolor{red}{red} tokens will be instantiated for patch generation via static analysis. In particular, \codeIn{EXPR[T]} denotes the type of expression \codeIn{EXPR} is constrained by \codeIn{T}.}
    \label{tab:table_abstract}
    \centering
    \resizebox{\textwidth}{!}{
    \begin{tabular}{l|l|l|l|l}
    \toprule
    & \multirow{1}{*}{\textbf{AST Node Type}} & \multirow{1}{*}{\textbf{Example}} & \multirow{1}{*}{\textbf{Skeleton}} & \textbf{Constraint Description}\\
    \midrule
     \multirow{11}{*}{\codeIn{STMT}} & \multirow{1}{*}{AssertStatement} & \multirow{1}{*}{\codeIn{assert a $>$ 0;}} & \multirow{1}{*}{\codeIn{\keyw{assert} EXPR[T];}} & \codeIn{T} is \codeIn{boolean}\\
    & \multirow{1}{*}{ConstructorInvocation} & \multirow{1}{*}{\codeIn{this(a);}} & \multirow{1}{*}{\codeIn{\keyw{this}(EXPR$_1$[T$_1$],...,EXPR$_n$[T$_n$]);}} & $n$ and \codeIn{T$_n$} should be compatible with the class.\\
    & \multirow{1}{*}{DoStatement} & \multirow{1}{*}{\codeIn{do\{ S \}while(a $<$ 0);}} & \multirow{1}{*}{\codeIn{\keyw{do}\{STMT\} \keyw{while}(EXPR[T]);}} & \codeIn{T} is \codeIn{boolean}\\
    & \multirow{1}{*}{ForStatement} & \multirow{1}{*}{\codeIn{for( ;a$<$0; )\{ S \}}} & \multirow{1}{*}{\codeIn{\keyw{for}(;EXPR[T];)\{STMT\}}} & \codeIn{T} is \codeIn{boolean}\\
    & \multirow{1}{*}{IfStatement} & \multirow{1}{*}{\codeIn{if(a $<$ 0)\{ S \}}} & \multirow{1}{*}{\codeIn{\keyw{if}(EXPR[T])\{STMT\}}} & \codeIn{T} is \codeIn{boolean}\\
     & \multirow{1}{*}{ReturnStatement} & \multirow{1}{*}{\codeIn{return a;}} & \multirow{1}{*}{\codeIn{\keyw{return} EXPR[T];}} & \codeIn{T} is compatible with the return type\\
     & \multirow{1}{*}{SwitchStatement} & \multirow{1}{*}{\codeIn{switch(a)\{case b: f();\} }} & \multirow{1}{*}{\codeIn{\keyw{switch}(EXPR1[T$_1$])\{\keyw{case} EXPR2[T$_2$]: STMT\} }} & \codeIn{T$_1$} and \codeIn{T$_2$} are compatible\\
    & \multirow{1}{*}{ThrowStatement} & \multirow{1}{*}{\codeIn{throw a;}} & \multirow{1}{*}{\codeIn{\keyw{throw} EXPR[T];}} & \codeIn{T} is \codeIn{Exception}\\
    & \multirow{1}{*}{VarDeclStatement} & \multirow{1}{*}{\codeIn{int a = b;}} & \multirow{1}{*}{\codeIn{\keyw{int} EXPR[T];}} & \codeIn{T} is compatible with \codeIn{int}\\
    & \multirow{1}{*}{WhileStatement} & \multirow{1}{*}{\codeIn{while(a $<$ 0)\{S\}}} & \multirow{1}{*}{\codeIn{\keyw{while}(EXPR[T])\{STMT\}}} & \codeIn{T} is \codeIn{boolean}\\
    & ExpressionStatement & \codeIn{a = a + b;}& \codeIn{EXPR[T];} & No constraint on \codeIn{T}\\
    \hline
    \multirow{14}{*}{\codeIn{EXPR[T$_0$]}} & \multirow{1}{*}{Assignment} & \multirow{1}{*}{\codeIn{a = a + b}} & \multirow{1}{*}{\codeIn{\highlight{VAR}[T$_1$]=EXPR[T$_2$]}} & \codeIn{T$_1$} and \codeIn{T$_2$} are compatible\\
     & \multirow{1}{*}{CastExpression} & \multirow{1}{*}{\codeIn{(int) b}} & \multirow{1}{*}{\codeIn{(\keyw{int}) EXPR[T]}} & \codeIn{T} is compatible with \codeIn{int}\\
    & \multirow{1}{*}{ClassInstanceCreation} & \multirow{1}{*}{\codeIn{new ClassA(a,b)}}  & \multirow{1}{*}{\codeIn{\keyw{new} \highlight{CNAME}(EXPR$_1$[T$_1$],...EXPR$_n$[T$_n$])}} & class \codeIn{CNAME} is compatible with \codeIn{T$_0$}; $n$ and \codeIn{T$_n$} fit \codeIn{CNAME}\\
    & \multirow{1}{*}{ConditionalExpression} & \multirow{1}{*}{\codeIn{a $>$ b ? a : b}} & \multirow{1}{*}{\codeIn{EXPR$_1$[T$_1$]?EXPR$_2$[T$_2$]:EXPR$_3$[T$_3$]}} & \codeIn{T$_1$} is \codeIn{boolean}; \codeIn{T$_2$}, \codeIn{T$_3$} are compatible with \codeIn{T$_0$}\\
    & \multirow{1}{*}{FieldAccess} & \multirow{1}{*}{\codeIn{a.b}} & \multirow{1}{*}{\codeIn{EXPR[T$_1$].\highlight{VAR}[T$_2$]}} & \codeIn{VAR} is defined in \codeIn{T$_1$}; \codeIn{T$_2$}, \codeIn{T$_0$} are compatible\\
    & \multirow{1}{*}{InfixExpression} & \multirow{1}{*}{\codeIn{a + b}} & \multirow{1}{*}{\codeIn{EXPR$_1$[T$_1$] \highlight{INFIX\_OP} EXPR$_2$[T$_2$]}} & \codeIn{T$_1$}, \codeIn{T$_2$} are compatible with \codeIn{INFIX\_OP}\\
    & \multirow{1}{*}{PrefixExpression} &\multirow{1}{*}{\codeIn{!a.isEmpty()}} & \multirow{1}{*}{\codeIn{\highlight{PREFIX\_OP} EXPR[T]}} & \codeIn{T}, \codeIn{PREFIX\_OP} are compatible\\
    & \multirow{1}{*}{PostfixExpression} &\multirow{1}{*}{\codeIn{a++}} & \multirow{1}{*}{\codeIn{EXPR[T] \highlight{POSTFIX\_OP}}} & \codeIn{T}, \codeIn{POSTFIX\_OP} are compatible\\
    & \multirow{1}{*}{MethodInvocation} & \multirow{1}{*}{\codeIn{a.method(b)}} & \multirow{1}{*}{\codeIn{EXPR.\highlight{FNAME}(EXPR$_1$[T$_1$],...,EXPR$_n$[T$_n$])}} & Return type of \codeIn{FNAME} is compatible with \codeIn{T$_0$}; $n$ and \codeIn{T$_n$} fit \codeIn{FNAME}\\
    & \multirow{1}{*}{SimpleName} & \multirow{1}{*}{\codeIn{a}} & \multirow{1}{*}{\codeIn{\highlight{VAR}[T]}} & \codeIn{T} and \codeIn{T$_0$} are compatible\\
    & \multirow{1}{*}{SuperFieldAccess} & \multirow{1}{*}{\codeIn{super.a}} & \multirow{1}{*}{\codeIn{\keyw{super}.EXPR[T]}} & \codeIn{T} and \codeIn{T$_0$} are compatible\\
    & \multirow{1}{*}{SuperMethodInvocation} &\multirow{1}{*}{\codeIn{super.a(b)}} & \multirow{1}{*}{\codeIn{\keyw{super}.\highlight{FNAME}(EXPR$_1$[T$_1$],...,EXPR$_n$[T$_n$])}} & Return type of \codeIn{FNAME} is compatible with \codeIn{T$_0$}; $n$ and \codeIn{T$_n$} fit \codeIn{FNAME}\\
    & \multirow{1}{*}{VarDeclExpression} & \multirow{1}{*}{\codeIn{int a = b}} & \multirow{1}{*}{\codeIn{\keyw{int} \keyw
a = EXPR[T]}} & \codeIn{T} is compatible with \codeIn{int}\\
    & \multirow{1}{*}{VarDeclFragment} & \multirow{1}{*}{\codeIn{a = b}} & \multirow{1}{*}{\codeIn{\keyw{a} = EXPR[T]}} & \codeIn{T} is compatible with \codeIn{T$_0$}\\
    \bottomrule
    \end{tabular}
    }
\end{table*}

More specifically, we have defined a set of code abstraction rules by following the AST node definition in the Java Development Toolkit~\cite{JDTAST}. Table~\ref{tab:table_abstract} presents the details of the rules. In the table, the first column presents the abstracted notations that can be further abstracted according to their actual AST node types as shown in the second column. That is, the abstraction process is an recursive process in a top-down fashion by following the abstract syntax tree structure of the code until it cannot be further abstracted by any rules, e.g., individual variables or operators. Particularly, we provide a simple example (the 3rd column) for each type of AST node for better understanding of the skeleton construction rules (the 4th column). The last column describes the constraints that have to be satisfied when instantiating the skeleton for patch generation. Taking the AssertStatement ``\codeIn{assert a>0;}'' as an example, the abstracted skeleton will be ``\codeIn{assert EXPR[boolean]}'', where the ``\codeIn{EXPR}'' will be further abstracted into ``\codeIn{VAR INFIX\_OP 0}'' according to the rule for InfixExpression. Please note that we do not abstract constant values (e.g., 0) in the code since they are not program-specific elements and thus can be reused directly. Consequently, the ultimate skeleton will be ``\codeIn{assert VAR INFIX\_OP 0;}'', where \codeIn{VAR} has to be a variable of number type (e.g., \codeIn{int} and \codeIn{float}) and the operator \codeIn{INFIX\_OP} has to be logical comparators (e.g., $>$ and $<$). In this way, the structure of the LLM-generated patches can be preserved for effectively constraining the patch space, and the abstracted tokens (colored \textcolor{red}{red}) can be instantiated via analyzing the contexts under repair for tailoring to certain programs.

\subsection{Patch Instantiation}
\label{sec:section3.2}

According to the constructed patch skeletons introduced above, the patch instantiation process becomes straightforward -- replacing the abstracted tokens in the skeleton with concrete program elements that satisfy the given constraints. Specifically, there are in total four types of abstracted tokens (see Table~\ref{tab:table_abstract}) in the ultimate skeletons for instantiation, i.e., variables (\codeIn{VAR}), classes (\codeIn{CNAME}), method calls (\codeIn{FNAME}), and operators (\codeIn{INFIX\_OP}, \codeIn{PREFIX\_OP}, and \codeIn{POSTFIX\_OP}). While the skeleton can already effectively constrain the patch space, randomly generating patches from them may still encounter a large search space. Therefore, it is less efficient and easy to fail due to a limited time budget. Therefore, \TOOL{} incorporates a context-aware patch generation strategy during skeleton instantiation, which involves three optimizations: 
\begin{itemize}[leftmargin=*]
    \item \textbf{Element selection:} All used elements have to be usable under certain contexts and meet the constraints of used patch skeletons. For operators, \TOOL{} enumerates all type-compatible ones belonging to each type during skeleton instantiation since the number of them is very small ($<$10), while for the other three types of tokens, \TOOL{} includes a static analysis process for collecting all usable ones. Specifically, for variables, \TOOL{} records their types and scopes; for classes, it records their inheritance relations and accessible fields; for method calls, it records their complete signatures, including the required arguments, return types and classes which they belong to. Such information will determine the usability of program elements by checking the constraints associated to the skeleton.
    \item \textbf{Context similarity:} \TOOL{} considers two kinds of similarities, one of which is between the instantiated patch and the buggy code while the other is between instantiated patch and the LLM-generated patch. The first similarity is inspired by previous studies~\cite{kolak2022patch, Xia2022LessTM, prenner2022can, jiang2023impact, xia2023automated,xia2023plastic,wei2023copiloting, zhang2023gamma, silva2023repairllama}, which reported that the desired patches in realistic scenarios often involve small code changes. The second similarity is inspired by our insights and the prominent results of LLMs as they can provide valuable material for patch generation. To achieve these goals, \TOOL{} first preserves the common code elements (e.g., variables) used in both the buggy code and the LLM patch as much as possible. For different elements, \TOOL{} prefers patches that are ``close'' to the LLM patch. Specifically, it uses the general token-level edit distance~\cite{ke2015repairing,koyuncu2020fixminer} to measure the closeness between patches.
    \item \textbf{Adaptive application:} As explained in Section~\ref{sec:motivation}, it is possible that only a part of the LLM patch is desirable. Therefore, if the patches with all code changes failed to repair the bug, \TOOL{} adaptively applies a subset of the extracted modifications. Specifically, \TOOL{} endeavors to apply at most three individual modifications in one patch (producing a reasonable and manageable number of patches), where it prefers to select the most complex modifications for combination since they can make the candidate patches ``closer'' to the LLM-generated patches as explained above.
    
\end{itemize}

Based on the above patch instantiation process, given an LLM-generated patch, \TOOL{} generates candidate patches on top of the abstracted patch skeletons, which can effectively constrain the search space of patches.

\subsection{Patch Ranking and Validation}
\label{sec:section3.3}

To make the most probable patches to be evaluated early, we have developed a patch ranking strategy. When given a set of candidate patches generated by LLMs, \TOOL{} favors those that can offer more new resources for patch generation. Specifically, \TOOL{} assesses the number of \textbf{Insert(*)} and \textbf{Update(*)} modifications involved in a patch since they can bring new program elements/structures while \textbf{Delete(*)} cannot. The higher the count of these modifications, the higher the rank of the patch. Subsequently, for each LLM-generated patch, \TOOL{} constructs candidate patches using the patch instantiation process introduced earlier. Finally, \TOOL{} executes the associated test cases after applying each patch and identifies the ones that pass all the test cases as plausible patches, in line with existing studies~\cite{zhu2023tare,xia2023plastic,wei2023copiloting, zhang2023gamma, Xia2022LessTM}. In this process, \TOOL{} utilizes the ExpressAPR~\cite{xiao2023expressapr} framework for managing the test running. Consistent with existing studies~\cite{zhu2023tare,xia2023plastic,wei2023copiloting, zhang2023gamma, Xia2022LessTM}, a patch is deemed correct only if it is semantically equivalent to the developer patch, as determined through manual inspection.

\section{Experimental Setup}
\label{sec:Experimental Setup}

\subsection{Research Questions}
\label{sec:rq}

In this paper, we aim to answer the following research questions for evaluating the effectiveness of \TOOL{}.

\begin{itemize}[leftmargin=*]
    \item \textbf{RQ1:}
    \textbf{\textit{How effective is \TOOL{} for improving LLMs in repairing real-world bugs?}} In this RQ, we explore whether \TOOL{} can improve the effectiveness of existing LLMs in the task of program repair. Specifically, we integrate \TOOL{} with different LLMs and check whether it can correctly repair more bugs than using the LLM-generated patches directly.  
    \item \textbf{RQ2:}
    \textbf{\textit{How effective is \TOOL{} compared to the state-of-the-art APR tools?}}
    In this RQ, we integrate \TOOL{} with existing LLMs to form a standalone APR tool by following existing study~\cite{xia2023plastic,wei2023copiloting,zhang2023gamma}, and then compare its performance with a set of state-of-the-art APRs.
    
    \item \textbf{RQ3:} 
    \textbf{\textit{What is the contribution of each component in \TOOL{}?}} As shown in Table~\ref{tab:table_abstract}, \TOOL{} includes a set of abstraction rules, which play a critical role. In this RQ, we study the contribution of each rule to the effectiveness of \TOOL{}, and investigate the reasons behind \TOOL{}'s capability to fix complex bugs.
\end{itemize}

\subsection{Subjects}
\label{sec:subject}


In our experiment, we employed the widely-studied Defects4J~\cite{just2014defects4j} benchmark. 
In particular, we adopted both version 1.2 and version 2.0 of the benchmark for evaluating the generality of \TOOL{}. Specifically, Defects4J v1.2 consists of 391 bugs from six real-world projects, while Defects4J v2.0 includes another 438 bugs from 11 real-world projects. By following existing studies~\cite{xia2023automated, jiang2023impact}, we leveraged LLMs to generate candidate patches when providing the buggy function. The reasons are twofold: (1) The code length of a single function is suitable for the input and output of current LLMs, and the entire method can offer local contexts for LLMs; (2) Function-level fault localization is more precise than the line-level fault localization, and thus the APRs depending on the former can be more practical for real use. Therefore, we removed the bugs that require cross-function modifications. Consequently, we use all 255 single-function bugs from Defects4J v1.2 and 228 single-function bugs from Defects4J v2.0 in our evaluation.



\begin{table}[tb]
    \centering
    \caption{Details of baseline APRs}
    \label{tab:baseline}
    \begin{tabular}{c|c|c|c|c|c}
    \toprule
        \textbf{APR} & \textbf{Year} & \textbf{Venue} & \textbf{APR} & \textbf{Year} & \textbf{Venue} \\
    \midrule
         Tare& 2023 & ICSE & TBar& 2019 & ISSTA \\
         KNOD& 2023 & ICSE & PraPR& 2019 & ASE \\
         FitRepair& 2023 & ASE & AVATAR & 2019 & SANER \\
         GAMMA& 2023 & ASE & SimFix& 2018 & ISSTA \\
         Repilot& 2023 & FSE & CapGen& 2018 & ICSE \\
         AlphaRepair& 2022 & FSE & SketchFix & 2018 & ICSE\\
         Hanabi& 2022 & TOSEM & JAID & 2017 & ASE \\
         SelfAPR& 2022 & ASE & jGenProg & 2017 & ESE\\
         Recoder& 2021 & FSE & jKali & 2016 & ISSTA \\
         CURE& 2021 & ICSE & jMutRepair& 2016 & ISSTA\\
         FixMiner& 2020 & ESE & NOPOL & 2016 & TSE\\
    \toprule
    \end{tabular}
    
\end{table}

\subsection{Baselines and Metrics}
\label{sec:baselines}

To answer RQ1, we selected four commonly-used LLMs as the baselines, including two general-purpose LLMs (GPT-3.5 Turbo~\cite{radford2019language,brown2020language} and Llama-2~\cite{touvron2023llama}) and two code-specific LLMs (StarCoder~\cite{li2023starcoder} and CodeLlama~\cite{roziere2023code}), all of which have been used in diverse software engineering tasks~\cite{schafer2023empirical, zheng2023survey} and demonstrated to be effective, including automated program repair~\cite{silva2023repairllama, xia2023keep}.

To answer RQ2, we mainly compared \TOOL{} with the \textbf{ten} latest and best-performing APRs due to the space limit, including \textit{four LLM-based} APRs (FitRepair~\cite{xia2023plastic}, Repilot~\cite{wei2023copiloting}, GAMMA~\cite{zhang2023gamma}, and AlphaRepair~\cite{Xia2022LessTM}), \textit{four specially-designed deep-learning-based} APRs (Tare~\cite{zhu2023tare}, CURE~\cite{jiang2021cure}, Recoder~\cite{zhu2021syntax}, and Hanabi~\cite{xiong2022l2s}), \textit{one template-based} APR (TBar~\cite{liu2019tbar}), and \textit{one heuristic-based} APR (SimFix~\cite{jiang2018shaping}). Furthermore, to investigate whether our approach can significantly advance the research of APR and repair unique bugs, we further compared \TOOL{} with the results of \textbf{22} diverse APRs. Besides the ten mentioned above, the other 12 APRs are  
KNOD~\cite{jiang2023knod}, SelfAPR~\cite{ye2022selfapr},
PraPR~\cite{ghanbari2019prapr}, AVATAR~\cite{liu2019avatar}, FixMiner~\cite{koyuncu2020fixminer}, CapGen~\cite{wen2018context}, JAID~\cite{chen2017contract}, SketchFix~\cite{hua2018towards}, NOPOL~\cite{xuan2016nopol}, jGenProg~\cite{martinez2017automatic}, jMutRepair~\cite{martinez2016astor}, and jKali~\cite{martinez2016astor}. We also present the details of these baselines in Table~\ref{tab:baseline}. 

In this paper, a patch is plausible if it can pass all the test cases, and a plausible patch is correct if it is semantically equivalent to the developer patch.
For result analysis, we mainly compare the number of bugs that can be correctly repaired by each baseline by following previous studies~\cite{zhu2023tare,xia2023plastic,wei2023copiloting, zhang2023gamma, Xia2022LessTM,jiang2021cure,zhu2021syntax, liu2019tbar, jiang2018shaping, xiong2022l2s}. Furthermore, we also compare their patch precision, which denotes the ratio of bugs with correct patches to the bugs with plausible patches.

\subsection{Implementation and Configuration}
\label{sec:impl}


\textbf{Baselines.} For LLMs, we used the models StarCoderBase (i.e., StarCoder-15.5B), CodeLlama-7B, Llama-2-13B, and GPT-3.5-turbo-0301 in our experiment. The first three models were downloaded from HuggingFace~\cite{HuggingFace2023} and then deployed on our local machines while the last GPT-3.5 online model was accessed via API requests~\cite{GPT35}. For each model, we reused the \textit{prompt} proposed by Xia et al.~\cite{xia2023automated}, and adopted the model default settings for patch generation -- Top-p Necleus Sampling~\cite{holtzman2019curious} with p = 0.95 and temperature = 0.8. For each bug, one LLM generates at most 200 patches. For other baseline APRs, we reused their experimental results from the corresponding publications directly.


\textbf{\TOOL{}.} We implemented \TOOL{} in Java with approximately 22k lines of code. During the repair process, \TOOL{} at most generates 500 candidate patches based on one patch skeleton from LLM-generated patches. Besides, following prior work~\cite{zhu2023tare,jiang2018shaping,liu2019tbar}, we offer a 5-hour time budget for repairing a single bug.

\textbf{Fault localization.} As mentioned in Section~\ref{sec:Introduction}, we comprehensively evaluated the performance of our approach under both perfect and imperfect (i.e., automated) fault localization. In the first scenario, we offered the LLMs the buggy function directly for patch generation. While in the second scenario, we followed prior studies~\cite{zhu2021syntax, liu2019tbar, jiang2018shaping} and employed the spectrum-based algorithm, Ochiai~\cite{abreu2006evaluation}, implemented in GZoltar~\cite{riboira2010gzoltar}, to obtain a list of functions. Then, following the function ranking, we tried to generate patches for each one until exceeding the time limit. Please note that some baseline APRs used a finer-gained line-level perfect fault localization (i.e., offering the buggy line) in their experiments, such as FitRepair and Repilot. Although it can be more accurate as it confines the patch space into a single line, we do not further unify this configuration in this paper because the baselines cannot work with the function-level fault localization. Nevertheless, this difference may underestimate the effectiveness of our approach when compared with the baselines.

\textbf{Experimental environment.} Our experiments were conducted on a local machine equipped with dual Intel Xeon 6388 CPUs, 512GB RAM, and four A800 GPUs, running Ubuntu 20.04.6LTS.

\section{Result Analysis}
\label{sec:Result Analysis}

\subsection{RQ1: Overall Effectiveness for Improving LLMs in APR}
\label{RQ:RQ1}

As explained in Section~\ref{sec:rq}, to evaluate whether our approach can better utilize the LLM-generated patches in program repair, we compare the results of \TOOL{} with the repair results when using the LLM-generated patches directly. Specifically, we selected four diverse LLMs for comparison, aiming to display the generality of our approach (see Section~\ref{sec:baselines}). 
In this experiment, we offer LLMs the function-level perfect fault localization by following existing studies~\cite{xia2023plastic, wei2023copiloting, Xia2022LessTM, xia2023automated, jiang2023impact,jiang2021cure}. Table~\ref{tab:table_1_improvements} presents the number of bugs that can be correctly repaired by each method. In the table, we use \TOOL{}$_{GPT-3.5}$ to represent the repair results when \TOOL{} takes the patches generated by GPT-3.5 as inputs.

\begin{table}[tb]
\centering
\caption{Result comparison with different LLMs}
\label{tab:table_1_improvements}
\resizebox{\columnwidth}{!}{
\begin{tabular}{l|cc|cc}
\toprule
\multirow{2}{*}{Tool} & \multicolumn{2}{c|}{Defects4J v1.2}  & \multicolumn{2}{c}{Defects4J v2.0}\\ 
\cline{2-5}
 & \#Correct & Impv.(\%) & \#Correct & Impv.(\%)\\
\midrule
GPT-3.5& 43 &  & 45 & \\
$\TOOL{}_{GPT-3.5}$ & 53 & 23.26 & 53 & 17.78\\
\midrule
StarCoder & 42 & & 44 & \\
$\TOOL{}_{StarCoder}$ & 55 & 30.95 & 54 & 22.73 \\
\midrule
CodeLlama & 40 &  & 34 &  \\
$\TOOL{}_{CodeLlama}$ & 51 & 27.50 & 43  & 26.47 \\
\midrule
Llama-2 & 19 &  & 18 &  \\
$\TOOL{}_{Llama-2}$ & 25 & 31.58 & 24 & 33.33\\
\midrule
Average$_{LLMs}$ & 36 &  &35.25 & \\
Average$_{\TOOL{}_{LLMs}}$ & \textbf{46} & \textbf{27.78} & \textbf{43.25} & \textbf{23.40}\\
\bottomrule
\end{tabular}
}
\end{table}

From the table, we can observe that \TOOL{} can effectively increase the number of correct fixes compared with using the LLM-generated patches directly. Specifically, on Defects4J v1.2, \TOOL{} increases the number of correct fixes from 43, 42, 40, 19 to 53, 55, 51, 25 respectively compared with the four LLMs. \textit{The relative improvement is up to 31.58\%, with an average increase of 27.78\%.} On Defects4J v2.0, \TOOL{} increases the number of correct fixes from 45, 44, 34, 18 to 53, 54, 43, 24. \textit{The relative improvement is up to 33.33\%, with an average increase of 23.40\%.} This results demonstrate the generalizability of \TOOL{} in enhancing repair performance of LLMs since it achieved relatively close effectiveness when comparing with diverse models over different benchmarks.

Furthermore, the repair performance of the selected LLMs in this paper is also consistent with prior work~\cite{zheng2023survey}: \textit{GPT-3.5$>$StarCoder$>$CodeLlama$>$Llama-2}. This reflects that training purposes tend to have a larger influence on the LLM's performance than LLM's sizes. For example, GPT-3.5, despite having a much larger model size than StarCoder, shows very close repair effectiveness in our experiment. In contrast, CodeLlama, which is fine-tuned from Llama-2 on code, demonstrates a significant improvement in patch generation. We leave the further exploration of this situation in a broader range to our future work.



\begin{figure}[htbp]
    \centering
    
    \includegraphics[width=\columnwidth]{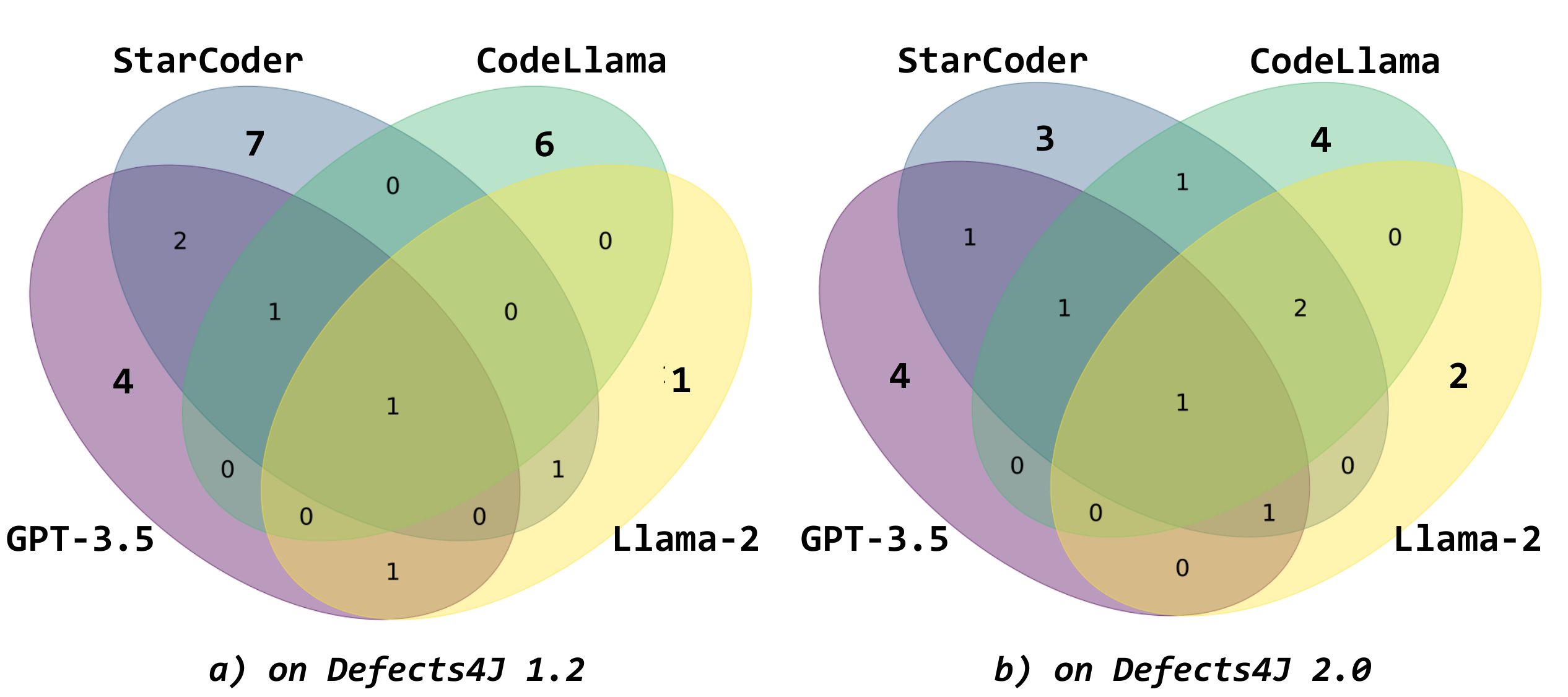}
    \caption{\TOOL{} uniquely repaired bugs when integrating with different LLMs.}
    \label{fig:venn_more_fixes}
\end{figure}

We also analyzed the complementary of different LLMs in this task. Figure~\ref{fig:venn_more_fixes} presents the number of bugs that are uniquely repaired when integrating \TOOL{} with each LLM. The results reveal that, although different LLMs performed diversely, they tend to complement each other as each individual LLM can offer the valuable repair guidance for some unique bugs. For example, Llama-2, which achieved the fewest correct fixes, contributed 3 unique fixes. This indicates that both code-specific and general-purpose LLMs should be considered in APR. In addition, the results also inspired us to explore whether \TOOL{} can still improve the repair performance when taking the complementary among different LLMs into consideration. Consequently, we combined the repair results of all the four LLMs, and then compared it with \TOOL{} that takes all their patches as inputs. The results show that the combination of the four LLMs successfully repaired 141 bugs on the two benchmarks, while \TOOL{} repaired 171 bugs (will be further discussed in Section~\ref{RQ2:RQ2}), demonstrating that \TOOL{} is indeed effective as it can effectively utilize the LLM-generated patches for better APR. In fact, when comparing with the latest GPT-4, \TOOL{} can still contribute unique fixes. We will discuss this result in Section~\ref{sec:gpt4}.


To show the necessity and effectiveness of our approach, we present two examples that cannot be repaired by LLMs but by \TOOL{}. Listing~\ref{lst:closure-36} and Listing~\ref{lst:cli-32} show the LLM-generated patches and the developer patches for two different bugs. From these examples we can observe that without sufficient program-specific knowledge (requiring \codeIn{getSingletonGetterClassName()} in Closure-36) and lacking of sufficient understanding of the contexts (assigning a value to \codeIn{pos}), LLMs struggle to generate the entire correct patch. Actually, correctly generating these two patches is challenging. The first patch needs to introduce a completely new \codeIn{if} statement from scratch, while the second patch deletes the whole \textit{while} loop (Deletion tends to produce incorrect patches and is usually disabled by existing APRs~\cite{jiang2018shaping,Tan2016anti}). However, empowered by the patch abstraction along with the context-aware and adaptive patch instantiation, \TOOL{} can effectively leverage the valuable guidance from LLMs' patches for patch generation.


\begin{lstlisting}[style=java, numbers=none,label=lst:closure-36,caption=Patch code of Closure-36 from Defects4J.]
//Developer patch
<@\hlc{green!15}{+\quad if (convention.getSingletonGetterClassName(callNode) != null) \{ }@>
<@\hlc{green!15}{+\qquad\quad return false; }@>
<@\hlc{green!15}{+\quad \} }@>
//LLM's patch
<@\hlc{yellow!15}{+\quad if (convention.getGetterMethod(callNode) != null) \{ }@>
<@\hlc{yellow!15}{+\qquad\quad return false; }@>
<@\hlc{yellow!15}{+\quad \} }@>
\end{lstlisting}

\begin{lstlisting}[style=java, numbers=none,label=lst:cli-32,caption=Patch code of Cli-32 from Defects4J.]
// Developer patch
<@\hlc{green!15}{- while ((pos <= text.length()) \&\& ((c = text.charAt(pos)) != ' ') \&\& (c != '\textbackslash n') \&\& (c != '\textbackslash r'))\{} @>
<@\hlc{green!15}{-\qquad ++pos; } @>
<@\hlc{green!15}{- \} } @>
// LLM's patch
<@\hlc{yellow!15}{- pos = startPos + width;} @>
<@\hlc{yellow!15}{- while ((pos <= text.length()) \&\& ((c = text.charAt(pos)) != ' ')  \&\& (c != '\textbackslash n') \&\& (c != '\textbackslash r')) \{} @>
<@\hlc{yellow!15}{-\qquad ++pos; } @>
<@\hlc{yellow!15}{- \} } @>
<@{\quad return pos == text.length() ? -1 : pos;} @>
\end{lstlisting}
\vspace{-6pt}

\subsection{RQ2: Effectiveness Compared with Baselines}
\label{RQ2:RQ2}
\begin{table*}[htbp]
    \caption{Repair results with perfect fault localization. In the table, GR represents our approach \TOOL{}.}
    \label{tab:table_2_compare}
    \centering
    \resizebox{\textwidth}{!}{
    \begin{tabular}{lc|cc|cc|ccccccccccc}
    \toprule
    \textbf{Project} & \textbf{\#Bugs}& \textbf{GR$_{\text{LLM}\times4}$} & \textbf{LLM$\times$4} & \textbf{GR$_{\text{LLM}\times2}$ }& \textbf{LLM$\times$2} & \textbf{FitRepair} & \textbf{Repilot}  & \textbf{Tare} & \textbf{GAMMA} & \textbf{AlphaRepair} & \textbf{CURE} &  \textbf{Recoder} & \textbf{TBar} \\
    \midrule
    Chart & 16 & 8 & 7 & 7 & 7 & 8 & 6 & 11 & 9 & 8 & 9 & 10 & 9 \\
    Closure & 93 & 32 & 20 & 25 & 17 & 29 & 21  & 22 & 20 & 22 & 13 & 20 & 15\\
    Lang & 42 & 14 & 11 & 8 & 7 & 17 & 15 & 13 & 10 & 11 & 9 & 10 & 10 \\
    Math & 72 & 26 & 23 & 25 & 21 & 23 & 20 & 20 & 19 & 19 & 16  & 16 & 16\\
    Time & 16 & 1 & 1 & 1 & 1 & 3 & 2  & 3 & 1 & 3 & 1 & 3 & 2 \\
    Mockito & 16 & 6 & 6 & 5 & 5 & 4 & 0 & 2 & 2 & 4 & 4 & 2 & 2 \\
    \midrule
    Defects4J v1.2 & 255 & \textbf{87} & 68 & 71 & 58 & 85 & 64 & 71 & 61 & 67 & 52 & 61 & 54 \\
    \midrule
    Defects4J v2.0 & 228 & \textbf{84} & 73 & 75 & 65 & 44 & 47 & 37 & 39 & 35 & 18  & - &  -\\
    \midrule
    Total & 483 & \textbf{171} & 141 & 146 & 121 & 129 & 111 & 108 & 100 & 102 & 70 & - & - \\
    \toprule
    \end{tabular}
    }
\end{table*}

\begin{table}[htbp]
    \footnotesize
    \caption{Repair results without perfect fault localization. X/Y denotes X correct patches and Y plausible patches.}
    \label{tab:table_3_SBFL}
    \centering
    \begin{tabular}{c|c|ccccc}
        \toprule
         \textbf{Project} & \textbf{\TOOL{}} & \textbf{Tare}  & \textbf{TBar} & \textbf{SimFix} & \textbf{Hanabi}\\
        \midrule 
        Chart & 7/10 & 11/14 & 7/10 & 4/5 & 1/3 \\
        Closure & 16/33 & 12/23 & 6/10 & 5/5 & -/- \\
        Lang & 12/19 & 12/19& 4/11 & 6/9 & 1/1 \\
        Math & 22/40 & 18/34 & 12/26 & 11/20 & 13/15 \\
        Time & 1/3 & 2/3 & 1/2 & 1/1 & 2/2 \\
        Mockito & 6/6 & 2/2 & 1/2 & -/- & -/- \\
        \midrule
        Total & \textbf{64/111} & 57/95 & 31/61 & 27/40 & 17/21 \\
        \midrule
        P(\%) & 57.66 & 60.00 & 50.82 & 67.50 & 80.95\\
        \bottomrule
    \end{tabular}
\end{table}


\textit{(1) Performance with perfect localization.} We compare our approach with the state-of-the-art APRs that were also evaluated under the assumption of perfect fault localization. As explained in Section~\ref{sec:impl}, \TOOL{} takes a buggy function as input and leverages LLMs to generate the initial patches, while the baseline APRs may take a buggy line as input.
Table~\ref{tab:table_2_compare} displays the number of bugs that can be correctly fixed by each APR on both Defects4J v1.2 and v2.0. We omitted the detailed results on Defects4J v2.0 due to space limit, which are available on our open-source repository. By following the baseline APR FitRepair~\cite{xia2023plastic}, \TOOL{} also takes the patches generated by four LLMs used in RQ1 as inputs. In addition, we also reported the performance of \TOOL{} when using two LLMs (i.e., GPT-3.5 and StarCoder) since it can be more resource-saving. The third and fifth columns present the corresponding results. We show the number of correct fixes by directly using the LLM-generated patches when combining the same LLMs as well. 

From the table, we can see that \TOOL{} can not only significantly improve the correct fixes compared with using the LLM-generated patches directly, it also significantly outperforms all baseline APRs. For example, compared with the best-performing FitRepair, \TOOL{} successfully repaired 42 more bugs. Specifically, \TOOL{} performs consistently well on different benchmarks. On the contrary, the baseline APRs tend to achieve better performance on Defects4J v1.2 than v2.0, indicating that our approach is more general and less likely to overfit certain benchmarks. As it will be presented in Section~\ref{leakage}, when using an another new benchmark that was never used by previous studies, \TOOL{} can still effectively repair a number of bugs.

\begin{figure}[htbp]
    \centering
        \includegraphics[width=\columnwidth]{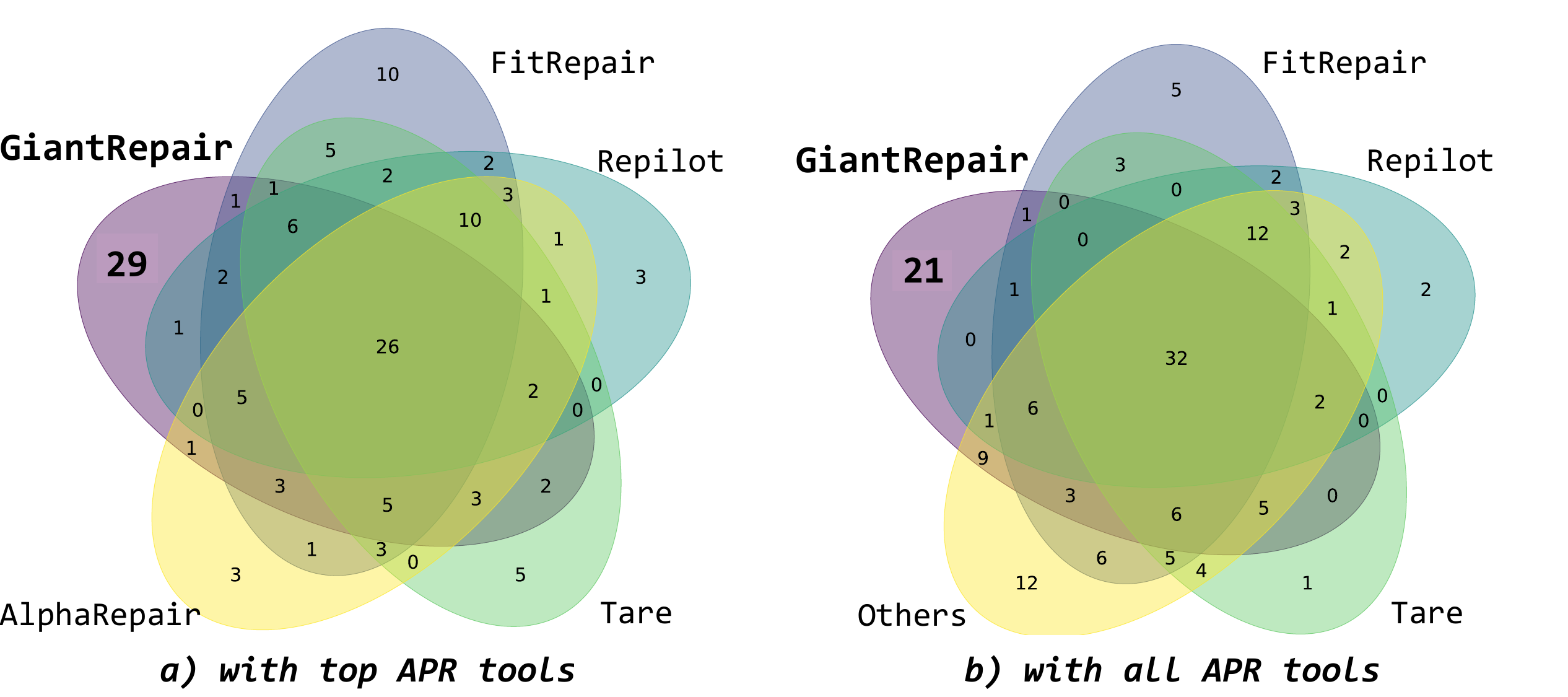}
    \caption{Uniquely repaired bugs on Defects4J v1.2}
    \label{fig:venn_baselines_12}
\end{figure}

Similarly, we further analyzed the complementary between our approach and the baseline APRs. In particular, since not all the 22 baselines were evaluated on Defects4J v2.0, we only compared their results on Defects4J v1.2. Figure~\ref{fig:venn_baselines_12} shows the overlaps of correct fixes by different APRs. Specifically, we first compare our approach with the top-4 baseline APRs (see Table~\ref{tab:table_2_compare}) in the left figure. Then, we compare our approach with all 22 baseline APRs (explained in Section~\ref{sec:baselines}) in the right figure. From the figures we can see that \TOOL{} has the capability to repair more new bugs compared with existing APRs. In particular, even compared with all 22 baselines, \TOOL{} can still repair 21 unique bugs, indicating its high effectiveness.
For instance, besides the two bugs introduced in Section~\ref{sec:motivation}, the bug shown in Listing~\ref{lst:closure-55} is an another example which can be correctly repaired by \TOOL{} but cannot by all the baseline APRs. To repair this bug, a new condition should be inserted. As a result, the patch skeleton (i.e., inserting a method call as the condition) generated from the LLM's patch can offer the valuable insights and effectively confine the search space.



In contrast, previous APR tools often struggled to find modifications similar to the correct fixes in the vast search space.
Additionally, many existing learning-based tools aim to only fix single-line bugs and are unable to address complex bugs requiring code changes across multiple lines.

\textit{(2) Performance with imperfect localization.} As mentioned in Section~\ref{sec:Introduction}, existing studies usually evaluated the performance of LLM-based APRs under the assumption of perfect fault localization. In this experiment, we further evaluate \TOOL{} in a more realistic application scenario with imperfect fault localization. Specifically, we used the automated fault localization results as introduced in Section~\ref{sec:impl}. Therefore, we compared it with baseline approaches that were also evaluated under the same setting. Table~\ref{tab:table_3_SBFL} shows the repair results of different methods on Defects4J v1.2 because all the baselines consistently used this benchmark. Moreover, we also report their patch precision like the baselines did. 

\begin{lstlisting}[style=java, numbers=none,label=lst:closure-55,caption=Patch code of Closure-55 from Defects4J.]
<@\lin{}{ static boolean isReduceableFunctionExpression(Node n) \{ }@>
// Developer patch
<@\hlc{green!15}{\lin{}\,+\qquad return NodeUtil.isFunctionExpression(n)} @>
<@\hlc{green!15}{\lin{}\,+\qquad\quad \&\& !NodeUtil.isGetOrSetKey(n.getParent());} @>
<@\hlc{green!15}{\lin{}\,- \qquad return NodeUtil.isFunctionExpression(n);}@>
// LLM's patch
<@\hlc{yellow!15}{\lin{}\,+\qquad return NodeUtil.isFunctionExpression(n)} @>
<@\hlc{yellow!15}{\lin{}\,+\qquad\quad \&\& !NodeUtil.isNameDeclaration(n.getParent());} @>
<@\hlc{yellow!15}{\lin{}\,- \qquad return NodeUtil.isFunctionExpression(n);}@>
<@\lin{}{ \}}@>
\end{lstlisting}

From the result we can conclude that \TOOL{} can also achieve better repair performance when using the imperfect fault localization than the best-performing APR. Specifically, \TOOL{} successfully repaired 64 bugs, which is even close to existing LLM-based APRs with perfect fault localization, such as Repilot and AlphaRepair. This is partially attributed to the reason that \TOOL{} relies on a coarse-grained function-level fault localization, which is much easier than that at the line level. However, the results also show that the powerful code generation ability of LLMs may also increase the risk of generating incorrect patches (i.e., low patch precision) due to the issue of weak tests~\cite{qi2015analysis,tian2022predicting,xiong2018identifying}. Nevertheless, the results indicate that \TOOL{} is able to be used in practice.

\begin{figure}[tbp]
    \centering
    \includegraphics[width=\columnwidth]{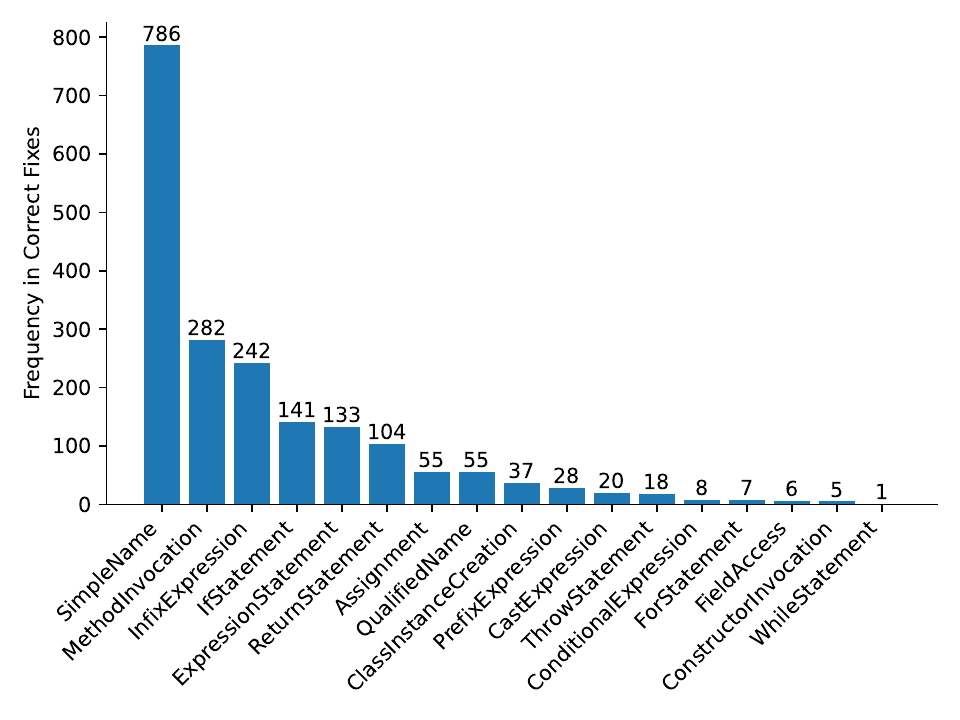}
    \caption{Component contribution of \TOOL{} }
    \label{fig:contribution}
\end{figure}

\subsection{RQ3: Contribution of Each Component in \TOOL{}}
\label{RQ3:RQ3}
In this section, we analyze the contributions of the skeleton abstraction rules shown in Table~\ref{tab:table_abstract} as they play a critical role in \TOOL{}. Figure~\ref{fig:contribution} shows the frequency of each rule that was involved in all the correct fixes of \TOOL{}.
Note that one fix may involve multiple same or different rules. For example, abstracting an \codeIn{if} statement may also need to abstract an infix expression in its condition. The \textit{x-axis} presents the associated AST node types of the rules. From the figure we can see that most rules contribute to the final correct fixes. In particular, {MethodInvocation}, {InfixExpression} and {IfStatement} are the most frequently used ones except for {SimpleName}, which aligns with existing conclusions~\cite{jiang2018shaping,liu2019tbar,saha2017elixir}. In particular, benefited from the patch skeleton, \TOOL{} is able to repair bugs that require complex modifications, such as inserting a completely new \codeIn{for} statement.
In summary, the abstraction rules in \TOOL{} are effective.

\section{Discussion}
\label{sec:diss}

\subsection{Comparing \TOOL{} with GPT-4}
\label{sec:gpt4}

As reported in RQ1, \TOOL{} can significantly increase the correct fixes compared with directly using the LLM-generated patches, including many unique fixes. However, since the repair ability of LLMs may also increase over the time. To investigate whether or not \TOOL{} is still effective for repairing unique bugs when comparing to the most advanced LLMs, we conducted another experiment with the online GPT-4-1106-preview~\cite{GPT4} (the latest version at the time of this study). Specifically, we randomly selected ten bugs that were correctly repaired by \TOOL{} but cannot by the studied LLMs, and then invoked GPT-4 via API requests to generate 20 patches for each bug. In this process, we used the same configuration and \textit{prompt} introduced in Section~\ref{sec:impl}. 
Our results show that GPT-4 only repaired one of the ten bugs, indicating that \TOOL{} can still useful even compared with the latest LLM. The detailed results of this experiment can be found at the our project webpage.

\subsection{Data leakage}
\label{leakage}

Data leakage is a common concern for LLM-based APRs. To investigate its impact on our conclusion, we first performed a manual analysis by following prior studies~\cite{Xia2022LessTM,wei2023copiloting, xia2023plastic}. Specifically, we analyzed whether the patched code has been used as the training data of the LLMs. We selected StarCoder as the representative as it is the only one that published its training data. The results show that among the 109 (=55+54) correct patches generated by \TOOL{}$_{StarCoder}$, 23 of them were included in StarCoder's training data. That is, a large majority of the correct patches 86/109 were not seen by StarCoder previously, and thus the effectiveness should come from the abilities of StarCoder and our patch generation method themselves, rather than the data leakage.

Moreover, to offer a more strong evidence of \TOOL{}'s effectiveness while avoiding data leakage, we conducted an extra experiment. We adopted a new benchmark -- GrowingBugs~\cite{GrowingBugsICSE21, GrowingBugsTSE2022,NaturalnessOfBugsFSE2022}. Specifically, we removed the projects that were involved in StarCoder's training data (34/250 projects left) and then filtered bugs that require cross-function modifications in these 34 projects (51/122 bugs left). Finally, we employed \TOOL{}$_{StarCoder}$ to repair them. The results show that it correctly repaired 10 out of the 51 bugs, further confirming the effectiveness of our approach.



\subsection{Limitation}
\label{sec:Limitation}
First, our experiments involved four LLMs (GPT-3.5-turbo, Llama-2, StarCoder and CodeLlama) and one programming language (Java). While this provides valuable insights, it still represents a limited scope in demonstrating the full capabilities of \TOOL{}, which is theoretically capable of utilizing any generative LLMs' generated patches in a wide range of programming languages. Another limitation of \TOOL{} lies in its time efficiency. The time LLMs take to generate patches is not accounted for in the patch generation process of \TOOL{}. Finally, \TOOL{} currently abstracts and instantiates code skeletons from single LLM-generated patch, while during our experiments, we found some bugs' correct fixes may be distributed across multiple patches. Investigating how to integrate these meaningful fixes from various patches could be a valuable focus for future development.



\subsection{Threats to validity}
\label{sec:Threats}
\textbf{Internal.} Manually reviewing all plausible patches to identify correct patches that are semantically consistent with the reference patch is an internal threat to the validity of our work. Following common APR practice, we perform a careful analysis of each plausible patch and have published our full set of correct and plausible patches. Another internal threat to validity is the LLMs used in our paper may trained on open-source code from GitHub, potentially overlapping with Defects4J dataset. To address this, we conduct a detailed discussion in Section~\ref{leakage} and employ a new dataset to demonstrate the effectiveness of \TOOL{}.

\textbf{External.} The primary external threat to validity comes from the evaluation datasets we used, and the performance of \TOOL{} may not be generalized to other datasets. To address this, we use two different datasets to evaluate \TOOL{}: Defects4J v1.2, Defects4J v2.0 and demonstrate that \TOOL{} is still effective and able to achieve state-of-the-art results. In the future, we plan to evaluate \TOOL{} on more datasets across multiple programming languages to address this threat.

\section{Related Work}
\label{sec:related}


Numerous APR approaches have been proposed to tackle the large search space of bugs, with the goal of enhancing the quality of generated patches. They employed a wide range of techniques, including predefined repair templates~\cite{martinez2016astor,long2017automatic,hua2018towards,ghanbari2019practical,jiang2019inferring,liu2019avatar}, heuristic rules~\cite{le2011genprog, Le2016HistoryDP,long2016automatic,xin2017leveraging,xiong2017precise,wen2018context,jiang2018shaping}, program synthesis ~\cite{xuan2016nopol,le2017s3,long2015staged,mechtaev2016angelix} and deep-learning ~\cite{chen2019sequencer,lutellier2020coconut,li2020dlfix, zhu2021syntax, jiang2021cure, ye2022neural, zhu2023tare}.

Recent works employed pre-trained models for APR tasks are more related. These approaches treat an APR task as a code generation task. AlphaRepair~\cite{Xia2022LessTM} masks out buggy code and uses CodeBERT~\cite{feng2020codebert} to replace the masked tokens with correct tokens to generate patches. GAMMA~\cite{zhang2023gamma} employs a similar process, applying masked fix patterns on buggy code and leveraging CodeBert and UniXcoder~\cite{guo2022unixcoder} to fill masked tokens. Different from these approaches, several studies~\cite{prenner2022can, xia2023automated, jiang2023impact} have explored the efficacy of directly applying LLMs to APR. Repilot~\cite{wei2023copiloting} fuse CodeT5~\cite{wang2021codet5} with a completion engine to improve performance.

Recent work, FitRepair~\cite{xia2023plastic} is the most closely related to \TOOL{} by employing the plastic surgery hypothesis in LLM-based APR to improve performance. FitRepair proposes two fine-tuning methods and one prompt format, utilizing four CodeT5~\cite{wang2021codet5} to generate patches. Different from FitRepair, \TOOL{} selects and abstracts the modifications in LLM-generated patches to generate patches. Consequently, \TOOL{} can fix more complex bugs.

\section{Conclusion}
\label{sec:Conclusion}
In this paper, we have proposed \TOOL{}, a novel automated program repair approach. Specifically, \TOOL{} leverages LLM-generated patches for patch skeleton construction and constraining the patch space, and then incorporates a context-aware skeleton instantiation process for generating high-quality patches tailored to specific programs. We have conducted two large-scale experiments for evaluating the effectivenes of \TOOL{}. The results demonstrated that it not only improved the correct fixes compared with pure LLMs, but also outperformed the latest state-of-the-art APRs.




\balance
\bibliographystyle{IEEEtran}
\bibliography{IEEEabrv,references}




\end{document}